# Mobiles as Portals for Interacting with Virtual Data Visualizations

**Michel Pahud**[1]
mpahud@microsoft.com

**Eyal Ofek**[1]
eyalofek@microsoft.com

**Nathalie Henry Riche**[1]
nath@microsoft.com

[1]Microsoft Research
One Microsoft Way
Redmond, WA, 98052

**Christophe Hurter**[2]
christophe.hurter@enac.fr

**Jens Grubert**[3]
jg@jensgrubert.de

[2]ENAC
Toulouse, France

[3]Coburg University of Applied
Sciences, Germany

## Abstract
We propose a set of techniques leveraging mobile devices as lenses to explore, interact and annotate n-dimensional data visualizations. The democratization of mobile devices, with their arrays of integrated sensors, opens up opportunities to create experiences for anyone to explore and interact with large information spaces anywhere. In this paper, we propose to revisit ideas behind the Chameleon prototype of Fitzmaurice et al. initially envisioned in the 90s [6, 7] for navigation, before spatially-aware devices became mainstream. We also take advantage of other input modalities such as pen and touch to not only navigate the space using the mobile as a lens, but interact and annotate it by adding toolglasses [3].

## ACM Classification Keywords
H.5.2. Information interfaces and presentation: (e.g., HCI): Input

## Introduction
The democratization of mobile devices, with their arrays of integrated sensors, opens up a large range of opportunities to create experiences for consuming and interacting with data visualizations. Mobile phones have a large audience and apply in a variety of context, from work to home scenarios, including on-the-go scenarios in public transports for example.

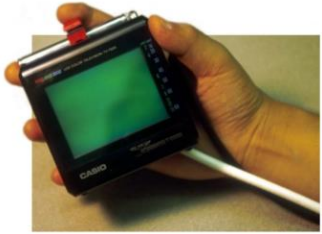

Figure 1: Palmtop prototype consisting of a handheld screen augmented with spatial sensors built in [6, 7].

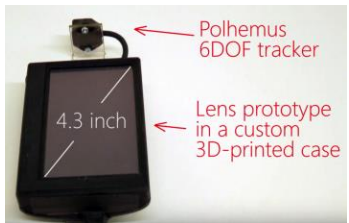

Figure 2: Chamaeleon lens prototype consisting of a touch screen and spatial sensors built in [16].

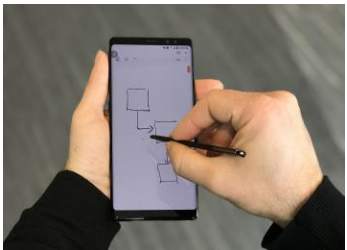

Figure 3: Today's mobiles available to the general public with a pen and touch screen and integrated sensors.

Yet, data visualization scenarios focused on visualizations of large and complex (multi-dimensional) data may benefit from more pixels than a small mobile screen can offer. In this paper, we explore the unique assets that mobile devices can offer and reflect on most compelling usage scenarios.

We propose to revisit ideas behind the Chameleon prototype [6, 7] for navigation, before spatially-aware devices became mainstream. We build on our previous work [16], a navigation technique using mobile handheld devices and expend it for the domain of data visualization. We also take advantage of other input modalities such as pen and touch to not only navigate the space using the mobile as a lens, but interact and annotate it by adding toolglasses [3].

## Spatially-Aware Devices

In 1993, Fitzmaurice et al. [6, 7] envisioned an experience for spatially-aware handheld devices. Spatially-aware devices are embedded with sensors which provides information on the location and orientation of the device. At that time, the prototype was built from a 4" LCD screen, camera, and a 6DOF tracker (Figure 1). About 20 years later, in 2013, we built upon this prototype and built a spatially-aware handheld touch screen [16] (Figure 2), still featuring an additional tracker. Since then, technologies evolved further, most mobile devices have integrated sensors (at least an accelerometer, a gyroscope, and often even a magnetometer) enabling to determinate their orientation as well as front and rear facing cameras allowing to detect their location. In addition, an ever larger number of devices feature touch and pen-enabled screens (Figure 3) making these spatially-aware handheld input devices available to the general public.

The particularity of the experiences envisioned in these previous work [6, 7, 16] is that they afford bimanual interaction: the user navigates the digital space (via pan and zoom) by holding and moving the device in the physical space with her non-preferred hand, and can interact with touch and a pen on the screen using her preferred hand.

## Views and Lenses for data visualization

There has been substantial research on the use of views and lenses in human-computer interaction and in data visualization. We point to surveys in this paper and give an overview of most relevant work. Views behaves as windows positioned over the visualization, enabling viewers to visualize data from different point of views [5]. Lenses are generally defined as views altering the content of a portion of an existing visualizations [21]. Many types of lenses have been explored, most related to our present work are tangible lenses using specific devices or mobiles and slates. For example, Spindler et al proposed a set of interaction techniques for the spatial navigation of data on a tabletop using tangible props in a spatial Augmented Reality setup [19]. They extended their work to the field of information visualization [20] and also employed user-perspective rendering for personalized views [18].

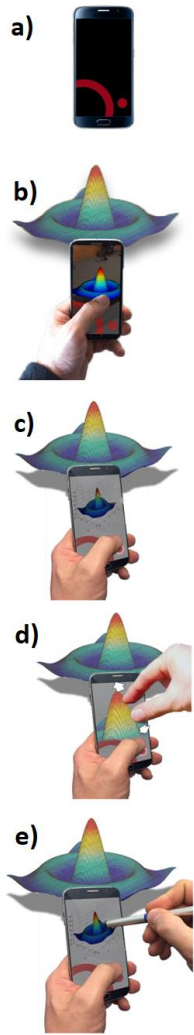

**Figure 4:** Interactions.

Further, researchers have investigated the use of multi-device ecologies for spatially-aware interaction with data visualizations, e.g., for interaction with map data using multiple smartphones and tablets localized through external positioning systems [17], using a combination of smartwatches and smartglasses [8]. Smartwatches, smartphones and large displays [13] or multiple self-tracked handheld devices for interaction with map or medical data [9]. Recently, researchers in information visualization have been exploring portals to view 3D data [2, 4].

## Portals for data visualization

In this paper, we propose to use a mobile phone as a portal on a virtual data visualization (Figure 3). Note that we opted to use the term portal as, in contrast to lenses, portals *do not augment* an existing display, and, in contrast to windows, portals enable the viewer to *interact* with the virtual object(s).

*Navigation*
In a nutshell, we leverage motion sensors to track the mobile in space and, thus, give the user the impression that the screen is a window on a digital object pinned at a given location in physical space and only visible through the mobile's screen (Figure 3). As users manipulate the mobile in space (moving and tilting it for example), corresponding views of the virtual visualization are updated on the screen to maintain the illusion. By interacting with buttons on the device and on the screen, people can interact with the virtual data visualization, select and annotation specific subsets of it for example.

*Clutching and Interactions*
Figure 4a shows our proposed user interface (for right handed users holding the device with the left hand) with a red arc used as thumb menu to gradually modify parameters such as information filter on the view, and a red dot used to clutch (or freeze) the view when desired. Moving the device in space allows the user to see the desired view of the data representation (Figure 4b) [6. 7]. When the user finds a view of interest, they can freeze the view by pressing a clutch button (Figure 4c). Once the view is clutched, the user can pinch to zoom or pan to precisely adjust the view if necessary (Figure 4d). The user could also use pen and/or touch to annotate it or select a region of interest by lassoing it (Figure 4e). Once the annotation is done, the user can release the clutch and continue to explore the data from different point of views. Alternatively, the user could use around-device gestures to spatially modify the 3D data visualization, [11], e.g., quickly rotating it or switching between viewing modes.

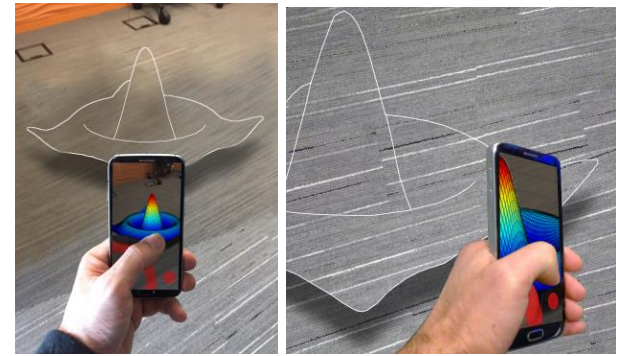

**Figure 3**: Our proposed mobile experience that would let the user view the data (e.g. 3D data) through the mobile device like a portal.

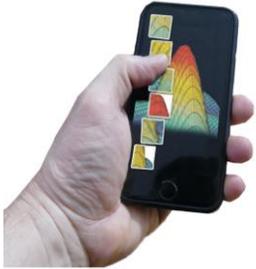

**Figure 5**: The user is accessing a set of bookmarks in the vicinities of the region he was exploring.

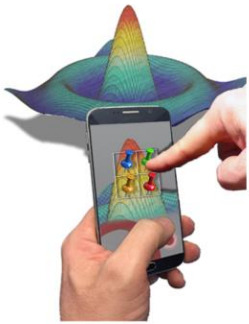

**Figure 6**: The user pinch to zoom and flick to pan on a view while the clutch is pressed. A toolglass allow the user to "push different color pins" as one would on a map to indicate different regions of interest.

*3D bookmarking*
When clutching a view and annotating it, the user may want to archive it for later retrieval. These bookmarks are then stored at their virtual location and can be later retrieved when the portal is in their vicinities. Selecting a bookmark (Figure 5) enables to retrieve the view as well as its annotations and continue the data exploration from this point. By sequencing and animating between a series of bookmarks, the user can author a tour that can then be shared with others.

*Toolglasses*
Leveraging the mobile screen can also enable to use semi-transparent toolglasses [3] to activate specific functions. For example, one could imagine a toolglass featuring different colored pins to indicate various regions of interest to revisit later (Figure 6).

## Usage scenarios

To illustrate the potential of portals for data visualization, we describe three different usage scenarios, highlighting novel experiences by constraining navigation for example or sculpting complex 3D selections.

1. **Non-Expert Planning a Hike**

As the technique effectively gives the illusion of manipulating a window over an invisible object, it requires little to no training and is therefore accessible to a large audience. We imagine that scenarios involving navigation of 3D models will be most compelling as previous work provides evidence that motion enables better comprehension of volumes [23].

For example, when planning a hike or preparing a flight plan, the portal can be used to get familiar with the terrain (3D virtual topological map) and its elevation, or in the case of a parasailing flight plan to get an overview of the wind currents in the area (by exploring a 3D wind model). Using their mobile while on-the-go, users will be able previewing the rest of the path as one would orient himself using map and compass. If users take photos and make notes during the hike, the experience can then be shared later by others who can explore the hike their friend did and access their notes and photos (in the same vein than panoramic 360 degree photo experiences shared on social media today).

*Constraint navigation*
As users plan their hike, one could imagine to overlay multiple hikes (trajectories) on the 3D terrain. Following a particular route may prove challenging as these routes may interweave and cross each other. By enabling the user to touch the screen to select a particular route, one could envision constraining the navigation in the virtual world (Figure 7). Once the navigation is constraint to follow a particular trajectory, motions of the portals are transferred to move forward or backward along it.

2. **Expert Executing Complex Selections**

The second scenario we envision is targeting expert analysts dealing with multidimensional data and composing complex queries. An example of multidimensional dataset that is naturally visualized as a 3D model are space time cube data visualizations [1] such as air traffic trajectories.

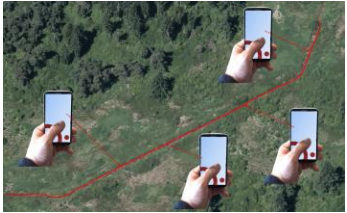

Figure 7: Lock on trajectory in 3D terrain illustration.

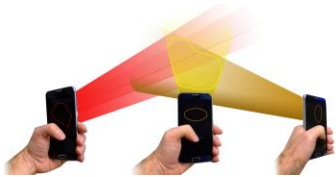

Figure 8: Selection by using different lasso shape, moving the device to get the intersection.

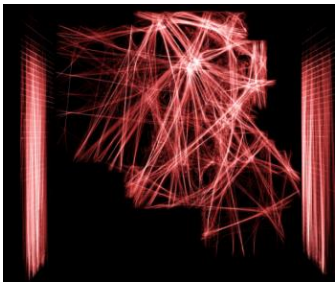

Figure 9: Air trajectory (space time cube).

Selecting subset of trajectories using a 2DOF input device such as the mouse is difficult, requiring multiple query refinement in different 2D cutting planes, which may be difficult for users to comprehend and execute accurately. Using a handheld device that can be directly placed in the 3D space has the potential to ease the composition of such complex 3D selections it is directly done in the 3D model space. Using their pen and moving their handheld device in the space, users can draw a 3D trajectory or shape. Using touch and pinch and zoom, they can edit each of its points as well as the radius of the cylinder, adjusting it to select trajectories within a certain range (Figure 8).

One can also envision that certain position in the space refer to different views of the data. For example, the faces of the bounding cube containing the data, can be used to depict 2D projections of the trajectories (e.g. looking downwards show a projection of all trajectory on x y coordinates revealing a heatmap of geographical flight routes).

### 3. Recording Tours for Storytelling

The third scenario focuses on storytelling. As scientific communication becomes more elaborate, experiences and videos depicting travel in space as a space ship orbits a planet[1] or following the path of a comet[2] immerse viewers and help them learn about science and technologies.

However, producing such videos require either dedicated multi-year development software[2] or a full multimedia department dedicated to produce these complex videos using professional 3D modeling and animation software. To enable a broader audience to create these video tours, we envision enabling them to experience 3D model with their phone device and enable them to record video shoots of the motion. The experience we envision is "what you see is what you record". Interacting on the screen using the dominant hand while the non-dominant hand is moving the device can also enable to add labels or pre-recorded annotations to their video. Ink and voice are also additional modalities that can be incorporated. As an example, Figure 9 should one of our early prototype that allowed to use the lens to navigate while tracing and recording voice, initially inspired by Wang FreeStyle[3].

The above scenarios were written with the intent to engage the discussion regarding future research around portals for data visualization. With the integration of motion sensors in many mobile devices today, we strongly believe that there are new research opportunities to explore the use of mobile as portals for interacting with data visualizations.

### Conclusion

This paper talk about prior art and our experience of building spatial aware device. It also shows the exciting possibilities with today's sensors which could make exploring data visualization very intuitive by moving a portal in space, interacting and annotating on the data using pen and touch.

---

[1] https://svs.gsfc.nasa.gov/12586

[2] http://www.worldwidetelescope.org

[3] https://www.youtube.com/watch?v=FRKzmFH7-cM